\documentclass[10pt,conference]{ieeetran}
\usepackage{amsmath}
\usepackage{mathptmx} % This is Times font
\usepackage{comment}
\usepackage{bm}
\usepackage{fancyhdr}
\usepackage[normalem]{ulem}
\usepackage[hyphens]{url}
\usepackage[sort,nocompress]{cite}
\usepackage[final]{microtype}
\usepackage[bookmarks=true,breaklinks=true,letterpaper=true,colorlinks,linkcolor=black,citecolor=blue,urlcolor=black]{hyperref}
\hypersetup{
    unicode=false,          % non-Latin characters in Acrobat’s bookmarks
    pdftoolbar=true,        % show Acrobat’s toolbar?
    pdfmenubar=true,        % show Acrobat’s menu?
    pdffitwindow=false,     % window fit to page when opened
    pdftitle={OCCAM},    % title
    pdfauthor={Mithuna Thottethodi},     % author
    pdfsubject={Occam: Optimal Data reuse for convolutional neural networks},   % subject of the document
    pdfcreator={Ashish Gondimalla},   % creator of the document
    pdfproducer={},         % producer of the document
    pdfkeywords={Accelerators,} {Convolutional neural networks,} % list of keywords
    pdfnewwindow=true,      % links in new window
    colorlinks=false,       % false: boxed links; true: colored links
    linkcolor=red,          % color of internal links
    citecolor=green,        % color of links to bibliography
    filecolor=magenta,      % color of file links=true,        
    urlcolor=cyan           % color of external links
}

\newtheorem{mydef}{Definition}

\fancypagestyle{firstpage}{
  \fancyhf{}
  
  \fancyfoot[C]{\thepage}
}

\title{Occam: Optimal Data Reuse for Convolutional Neural Networks\thanks{A previous version of this work was submitted in November 2017 to ISCA 2018.}
}
\author{\IEEEauthorblockN{Ashish Gondimalla}
\IEEEauthorblockA{\textit{Purdue University} \\}
\and
\IEEEauthorblockN{Jianqiao Liu}
\IEEEauthorblockA{\textit{Purdue University} \\}
\and
\IEEEauthorblockN{T.N. Vijaykumar}
\IEEEauthorblockA{\textit{Purdue University} \\}
\and
\IEEEauthorblockN{Mithuna Thottethodi}
\IEEEauthorblockA{\textit{Purdue University} \\}
}
\usepackage{graphicx}
\usepackage{ifthen}
\usepackage{calc}
\usepackage{pifont}
\usepackage{color}
\usepackage{fancyhdr}
\usepackage{xspace}
\usepackage{multirow}
\usepackage[table]{xcolor} 
\definecolor{lightgray}{gray}{0.9}
\newenvironment{stripetabular}{\rowcolors{2}{white}{lightgray}\tabular}{\endtabular}

\newcounter{hours}
\newcounter{minutes}

\newboolean{timeofmake}
\setboolean{timeofmake}{false}

\newcommand{\ignore}[1]{}

\newcommand{\mithuna}[1]{{\color{black} #1}\xspace} %Changing color to black for highlighted text

\newcommand{\dontinclude}[1]{ }

\newcommand{\eql}[1]{\label{eq:#1}}

\newcommand{\putsec}[2]{\vspace{-0.05in}\section{#2}\label{sec:#1}\vspace{-0.01in}}
\newcommand{\putsubsec}[2]{\vspace{-0.05in}\subsection{#2}\label{sec:#1}\vspace{-0.01in}}
\newcommand{\putsubsubsec}[2]{\vspace{0.05in}\subsubsection{#2}\label{sec:#1}\vspace{0.01in}}

\newcommand{\tabput}[3]{
\begin{table}[t]
\small %also for forcing a baselinestretch update
\begin{center}
{
#2
}
\end{center}
\caption{\footnotesize\sf #3 \label{tab:#1}}
\vspace{-0.3in}
\end{table}
}

\newcommand{\tabputW}[3]{
\begin{table*}[t]
\begin{center}
{
#2
}
\end{center}
\caption{#3 \label{tab:#1}}
\vspace{-0.3in}
\end{table*}
}

\newcommand{\figput}[4][1.0\linewidth]{
\begin{figure}[t]
\begin{minipage}{\linewidth}
\footnotesize 
\begin{center}
\includegraphics[clip, width=#1]{figures/#2}
\end{center}
\vspace{-0.15in}
\caption{#4 \label{fig:#2}}
\vspace{-0.2in}
\end{minipage}
\end{figure}
}

\newcommand{\figputW}[4][1.0\linewidth]{
\begin{figure*}[t]
\begin{minipage}{\linewidth}
\footnotesize 
\begin{center}
\includegraphics[clip, width=#1]{figures/#2}
\end{center}
\vspace{-0.2in}
\caption{#4 \label{fig:#2}}
\vspace{-0.2in}
\end{minipage}
\end{figure*}
}

\newcommand{\figref}[1]{Figure~\ref{fig:#1}}
\newcommand{\tabref}[1]{Table~\ref{tab:#1}}
\newcommand{\secref}[1]{Section~\ref{sec:#1}}

\IEEEoverridecommandlockouts

\begin{document}
\maketitle
\thispagestyle{firstpage}
\pagestyle{plain}

\begin{abstract}

Convolutional neural networks (CNNs) are emerging as powerful tools for image processing in important commercial applications. We focus on the important problem of improving the latency of image recognition.   
While CNNs are amenable highly to prefetching and multithreading to avoid memory latency issues, CNNs' large data -- each layer's input, filters, and output -- poses a memory bandwidth problem. 
While previous work captures only some of the enormous data reuse,  {\em full reuse} implies 
 that the initial input image and filters are read  once from off chip and the final output is written once off chip without spilling the intermediate layers' data to off-chip.  We propose 
{\em Occam} to capture full reuse via four contributions. First, we identify the necessary  condition for full reuse. Second,  we identify the   {\em dependence closure} as the sufficient condition to capture full reuse using the least on-chip memory. Third, because  the  dependence closure is often too large to fit in on-chip memory,  we propose a dynamic programming algorithm that optimally partitions a given  CNN 
to guarantee the least off-chip traffic at the partition boundaries
for a given on-chip capacity. While tiling is well-known, our contribution is determining the  optimal cross-layer tiles. Occam's partitions reside on different chips  forming a pipeline so that  a partition's filters and dependence closure remain on-chip as different images pass through (i.e., each partition incurs  off-chip traffic only for its inputs and outputs).
Finally, because the optimal partitions may result in an unbalanced pipeline, we propose {\em staggered asynchronous pipelines (STAP)} which  replicates the bottleneck stages  to improve throughput by staggering the mini-batches across the replicas.
Importantly, STAP achieves balanced pipelines {\em without} changing Occam's optimal partitioning. 
Our simulations show that, on average, Occam cuts off-chip transfers by 21x  and achieves  2.06x
and 1.36x better performance, and 33\% and 24\% better energy than  the base case and Layer Fusion, respectively. Using an FPGA implementation, Occam performs 5.1x better, on average, than the base case.  
 
\end{abstract}

\putsec{intro}{Introduction}

Advances in convolutional neural networks (CNNs)~\cite{lecun,alexnet,vggnet,resnet} have resulted in highly-accurate image classification and recognition. 
Current commercial applications are speech-based (e.g., voice assistants) which employ other types of neural networks (e.g., Long Short-Term Memory networks), so that CNNs currently
contribute only 5-10\% to machine learning workloads~\cite{TPU}. However,
emerging image-based applications (e.g., self-driving cars)  are poised to increase this contribution, as anticipated by several commercial CNN architectures (e.g., Movidius, Nervana, Nvidia). 

A CNN employs  numerous {\em filters} to identify features which are combined into higher-order features so that each CNN layer  applies the current layer's filters  to the previour layer's output features and outputs the next set of features, and eventually the final classification output.  Each layer applies a set of filters to its  {\em input feature map} by convolving each filter with the input map in two dimensions  -- i.e., ``sliding'' the filter along the input's two dimensions --  to extract the corresponding feature. Each layer simply puts together its filter results as its {\em output map}. The {\em weights} in the  filters are trained by reducing the error for the training inputs via back propagation.  
Like most previous CNN work, this paper focuses on the recognition phase. Further, this paper targets reducing the latency of recognition
as opposed to improving the   throughput. 
The latency goal is important for many interactive scenarios (e.g., self-driving cars) where trading-off latency for throughput is unacceptable~\cite{TPU}. 

The large number of layers (e.g., 34 in {\em resnet}) and filters per layers (e.g., 192), and large images (e.g., 224 x 224), result in  heavy compute and large intermediate data. The heavy  compute is somewhat offset by the abundance of regular parallelism amenable to GPGPUs~\cite{alexnet},  FPGAs~\cite{snowflake,ferdman-fpga2},  and TPUs~\cite{TPU}. Recent work  prunes the compute by using 8-bit fixed-point representation~\cite{quantization} or by exploiting zeroes in the data~\cite{moshovos-many,moshovos-cnn2}. TPU optimizes the compute via a novel systolic multiply-add array. However, the large intermediate  data continues to be a challenge for performance especially because more training data achieves higher accuracy but needs larger networks to avoid over-fitting~\cite{cnn-survey}.  Because we target data reuse, we describe CNNs in terms of the access patterns instead of neurons and synapses.   {\em Because CNNs are amenable to prefetching and multithreading, the problem is memory bandwidth
and not memory latency~\cite{TPU}.}

In the convolution,  as the filters ``slide over'' the input map, each input cell participates in many computations  which are repeated for each filter.  Each weight cell in a filter is reused by each input map cell.  Further, each layer's output map is the next layer's input map.  All of this reuse is for one input image and only in the convolution layers whereas other, fully-connected (FC) layers, do not have any reuse. The former account for more than 85\% of execution time in earlier CNNs~\cite{alexnet} and hence our focus, the latter 
are used only as the last layer in recent CNNs (e.g., {\em GoogLeNet}). 

Current practice is to write off-chip each layer's output map which is read back in by the next layer, losing inter-layer reuse. 
DianNao~\cite{diannao} and its successors~\cite{dadiannao,pudiannao,shidiannao} propose to reduce the off-chip traffic by placing all of the intermediate data and filters in an on-chip eDRAM, which may be inefficient (e.g., 50 and 21 MB for {\em VGGnet} and {\em Resnet}).    Other work propose to compress the data for the FC~\cite{eie} and convolutional layers~\cite{moshovos-cnn2,Dally-isca17}. 
While such compression reduces both the compute and memory volumes, 
the all-or-nothing approach works well only if the compressed data fits in on-chip memory and otherwise, generates off-chip traffic (e.g., recent CNNs may require several MBs even after compression). 
Further, such compression destroys compute and data access regularity hurting efficiency of GPGPUs, TPU and  FPGAs 
(SCNN employs crossbars for the irregularity).   

Exploiting reuse with reasonable on-chip memory to reduce the off-chip traffic is challenging.  Capturing  {\em full reuse} implies that the initial input image and filters are read  once from off chip and the final output is written once off chip without spilling the intermediate layers' data to off-chip. 
Because the filters  have high reuse, they are held on-chip (e.g., Eyeriss~\cite{eyeriss}), similar to our on-chip residence though the architectures hold at a time only one layer's filters  and reload each layer's filters once per input image. 
Further, capturing inter-layer reuse requires holding on-chip a layer's  output map to be read by the next layer. An output map cell depends on many input map cells -- the output cell's {\em dependence parents}, due to a dot product sub-computation in each convolution; and  many output map cells share the same dependence parents, which provides reuse. 
These output maps' dependence ancestors transitively extend to include the corresponding input maps of the earlier layers.  Accordingly, Layer Fusion~\cite{fusion}, a pioneering work, holds only the ancestors of an output map tile to capture some of the inter-layer reuse. While conventional tiling holds only the parents, this cross-layer tiling holds some or all of the ancestors similar to other work~\cite{polymage,halide} (\secref{related}). 

Despite these significant advances, none of the papers captures  full reuse.
To that end, we propose {\em Occam} and make the following contributions:

First, the {\em necessary condition} to exploit  full  reuse with the optimum (smallest) amount of on-chip memory is to hold  {\em one full input map row} or column, whichever is shorter. Due to convolution, each input map cell is a dependence parent of many output map cells {\em along both dimensions}. Capturing such two-dimensional reuse requires holding the full shorter dimension. 
Thus, the necessary condition determines the optimal tile {\em shape}. Unlike {\em matrix multiply}, a canonical tiling candidate,  the tile shape matters for CNNs (\secref{necessary}). 
Layer Fusion's tile, derived from another work~\cite{cong}, does not satisfy this condition incurring expensive recomputation triggered by reuse not captured on-chip. 

Second, Occam achieves full reuse by satisfying  the {\em sufficient condition} of holding  only the ancestors of a full output row or column all the way through all the layers to the initial input image -- i.e., the transitive closure of the dependence ancestors, called {\em the dependence closure} ({\em receptive field} in machine learning).  Layer Fusion identifies the closure but not for the full output row (the necessary condition).

Third, because a CNN's full dependence closure is often too large to fit in one chip's memory, we partition the CNN into sets of  contiguous layers so that each partition's dependence closure  fits  on-chip (each partition reads its input map from  and writes its output map to off chip).   Layer Fusion  chooses sub-optimal partitions because its exhaustive search is infeasible for  large networks  (e.g., $2^{34}$ choices in {\em Resnet}). Instead, we propose a dynamic programming (DP)  algorithm that optimally partitions a given  CNN 
to guarantee the least off-chip traffic at the partition boundaries
for a given on-chip memory capacity.
{\em While tiling is well-known, our contribution is determining the optimal cross-layer tiles.}
Fortunately, Occam's optimal partitions and tiles preserve CNNs' regular parallelism unlike prior compression work. 

Finally, the input-stationary approach~\cite{eyeriss}   holds on-chip the input/output maps and fetches the filters from off-chip, ignoring filter reuse across images (e.g., TPU). 
Occam's partitions reside on different chips  (e.g., GPU, TPU, or FPGA),  forming a pipeline so that  a partition's filters and dependence closure remain on-chip as different images pass through (i.e., each partition incurs  off-chip traffic only for its inputs and outputs). Occam  amortizes  filter loading to asymptotically zero cost over numerous images  to achieve {\em full cross-image reuse}.
While BrainWave~\cite{brainwave} holds the filters on-chip, its partitions are ad hoc, do not employ tiling, and  may incur pipeline imbalance.  
Because Occam's optimal partitions may also result in an unbalanced pipeline, we propose {\em staggered asynchronous pipelines (STAP)} which  replicates the bottleneck stages  to improve throughput by staggering the mini-batches across the replicas.
Importantly, STAP achieves balanced pipelines {\em without} changing Occam's optimal partitioning.

DP is a widely-used {\em meta} algorithmic method for diverse problems.
While our {\em optimal} DP formulation targets on-chip {\em memory},  another,  {\em heuristic} formulation~\cite{ferdman-fpga2} targets FPGA {\em compute} resources for CNNs. Similarly, PipeDream~\cite{pipedream-poster,pipedream-arxiv} employs DP to minimize  {\em pipeline imbalance in training} for a {\em given pipeline depth} without any data reuse considerations whereas Occam optimizes {\em data  reuse in inference} for a {\em given cache capacity} without constraining the pipeline depth. Further, because the large  data is problematic for any CNN architecture, such as GPGPU, TPU, or FPGA-based,  Occam is applicable to all the architectures. 
 
Our simulations show that, on average, Occam cuts off-chip transfers by 21x  and achieves  2.06x
and 1.36x better performance, and 33\% and 24\% better energy than  the base case and Layer Fusion, respectively. Using an FPGA implementation, Occam achieves 5.1x better average performance than the base case.

\putsec{background}{Reuse in CNN}

As mentioned in~\secref{intro}, a CNN comprises of many layers each of which employs many filters to extract higher-order features. A layer's output map is the next layer's input map. 
In general, each layer's input map is a cuboid whose height $h$,  width $w$, and channel count $m$  are analogous to the input image's height and width and three colors ({\em R}, {\em G}, and {\em B}); $m$ is the number of filters in the  previous layer (\figref{cnn}).

Each filter is another cuboid whose height  and  width are typically equal, $k$, and the channel count is the same as the input map's, $m$ (typically, $k << h, w$) (\figref{cnn}).  Each layer's output map dimensions are $h$ x $w$ x $n$, where  $n$ is the  number of filters in the layer. 
Some layers, called {\em pooling layers}, shrink $h$ and $w$ without changing $n$ by summarizing a set of  cells (e.g., maximum of four neighboring cells).  {\em A cell is a $1$ x $1$ x $1$ slice of a cuboid -- a scalar.
}

\figput{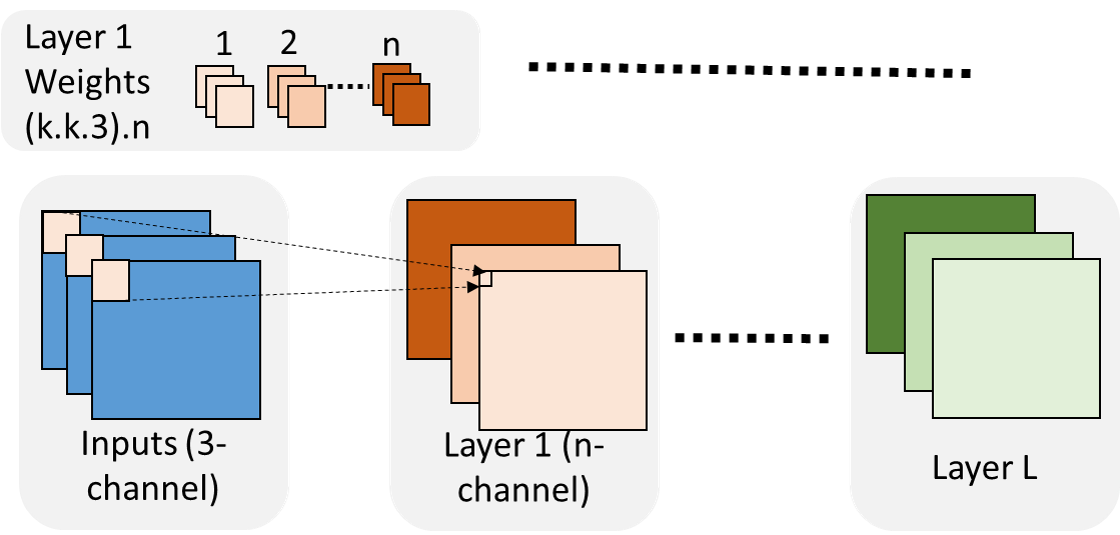}{}{Convolutional Neural Network}

Each output map cell is computed by a {\em cuboidal} product of the
input map and a filter (see the faint arrows from the Inputs to Layer 1 in~\figref{cnn}).  The next output map cell results from the  product with the filter ``slid over'' by a stride along the height and  width dimensions, one at a time,  in a two-dimensional convolution. 
There is no sliding along the third dimension of channels. 
Thus, each  cell in a layer's input map, except for those at the boundaries, is  reused $k *k$ times by a  filter in its various slid-over positions 
(assuming the convolution stride is $1$); this reuse is called {\em convolutional reuse}.  With $n$ filters in a layer, each input map cell produces $n$ output cells in the output channel dimension (\figref{cnn}). Thus, each input map cell is reused by an additional factor of $n$
called {\em cross-filter reuse}, for a total reuse of 
$k *  k * n$ times (e.g., for a layer with 128 3 x 3 filters, there is 1152 times reuse).  
Further, each layer's output map is the next layer's input map, providing another instance of reuse, called {\em inter-layer reuse}, for a total reuse of $k*k*n+2$ times. 

Past implementations use {\tt im2col} to transform the convolution into  a {\em matrix multiply}. To that end, the implementations replicate the input map cells for every slide position of the filters so that each filter can be matrix-multiplied with
an input-map tile of the same size. However, the replication causes enormous memory 
bloat ($k$-by-$k$ filters cause $k^2$ redundancy).Therefore, our and other 
recent implementations do not take this approach. 
However, there may be some replication in the private L1 caches due to parallel execution.  

\putsubsec{filter-reuse}{Reuse in filters}

A layer's filters are used only in that layer. Because each weight cell in a filter is ``slid over'' every input map cell in the height and width dimensions, each weight cell is reused $h * w$ times, ignoring boundaries (typically, {\em h}, {\em w} $>>$  {\em k}). To capture this reuse, current practice applies one filter at a time to the entire input map, refetching the input map, if too large to be held on chip, for every filter. In addition to capturing the filter reuse, this approach also  captures the $k*k$ times  reuse of the input map.  However, the approach does not capture all the reuse in the input map, requiring $(n + 2)*l$ refetches in case of large input maps
for  $l$ layers ($l$ is large for deep networks -- e.g., 34 in {\em ResNet}). Here, we make the simplifying assumption that all the layers' input maps and output maps are the same size. In~\secref{results}, we show results for real CNNs where this assumption is not true. 

\putsubsec{input-reuse}{Reuse in input map}

One way to avoid refetching the input map is by applying all the filters to each part of the input map before processing the next part. While this strategy spreads apart each filter's reuse (across input map parts), many recent CNN architectures instead hold the filters on chip because the filters are often larger than the input maps in later layers. However, due to their layer-by-layer processing these architectures hold  only one layer's filters at a time  so that 
each layer's filters have to be refetched  for the next image (i.e., no cross-image reuse as captured by Occam). 
For example, as noted in~\secref{intro}, Eyeriss~\cite{eyeriss} applies all the filters to each input map cell before the next cell capturing both convolutional reuse and cross-filter reuse -- a total  of $k*k*n$ times -- but not inter-layer reuse, resulting in  $2*l$ input map refetches for $l$ layers. {\em Residual} CNNs, in which a layer may get an additional input map from a previous layer~\cite{resnet}, require slightly more refetches: $2*l+r$ for $l$ layers of which $r$ have one residual input. 
Recall from~\secref{intro} that {\em full cross-image reuse} implies that the  initial input (final output) are read (written) once from (to) off-chip without spilling the intermediate layers' data. 
To capture inter-layer reuse, and thus  full reuse, we propose Occam.

Apart from convolution, CNN layers perform a few other computations, such as {\em batch normalization}, {\em pooling}, and {\em rectified linear unit (ReLU)}. These local operations are performed as the output map is produced and do not significantly change CNNs' reuse behavior.

\putsec{occam}{Occam}

Recall from~\secref{intro} that Occam makes 
% we propose Occam to capture full reuse via
four contributions. First, we specify the necessary condition for {\em full reuse} in terms of the tile shape. Second,  we propose an approach called {\em Dependence Closure} which satisfies the sufficient condition to capture full reuse using the least on-chip memory. Third, because the dependence closure  for full reuse is often too large,  we propose a dynamic programming algorithm that optimally partitions a given CNN while guaranteeing the least off-chip traffic at the partition boundaries for a given  on-chip memory capacity. By holding the filters on-chip, our partitions achieve  {\em full cross-image reuse}. Finally, we propose {\em staggered asynchronous pipelining (STAP)} to balance the pipeline formed by the partitions. 

\putsubsec{necessary}{\mithuna{Necessary condition}}  

Due to convolutional reuse, each input map cell of a layer  is a {\em dependence parent} of many output map cells {\em along both dimensions}. For  full reuse with rectangular filters\footnote{We discuss non-rectangular filters at the end of this subsection.}, we need to hold on-chip at least one {\em full input map row-plane} or column-plane, whichever is smaller. {\em A row-plane corresponds to a cuboid of dimensions 1 full row x 1 x  n, where n is the number of channels.} 
This condition identifies the {\em tile shape} necessary for full reuse.

To prove this condition, we make the following three simplifying assumptions (removed later): (1) a single input channel, (2)  the input tile includes the element at $(0,0)$, and (3) a rectangular tile of dimensions $X_t \times Y_t$ from the input feature map of dimensions $X_{max} \times Y_{max}$.
To minimize capacity misses, the {\em largest} tile that fits in the cache, called {\em maximal tile}, is used. 

\noindent {\bf Proof sketch: }
To prove by contradiction, we assume that the input tile of dimensions $X_t \times Y_t$ does {\em not} span a full row (or column)  (i.e.,  $X_t < X_{max}$ and $Y_t < Y_{max}$) and show that full reuse is impossible.  \figref{necessary}(a) shows the initial tile position. 
Because of the two-dimensional sliding nature of convolutional reuse, some elements at the boundaries of the input tile are dependence parents for output cells that cannot be computed fully with the current input tile data (red squares indicating future reuse). Moreover, such future reuse exists along both the $X$ and $Y$ dimensions. 

We now  show that {\em any} tile movement -- {\em even a single step in either the $X$ or $Y$ dimensions} -- rules out full reuse. Without loss of generality, a tile movement in the $X$-dimension by some positive stride $d$ (integer $d>0$) means the corners of the tile are now at $(d,0), (d,Y_t), (X_t+d,0),$ and  $(X_t+d, Y_t)$. 
Because of maximal tiles, all the elements in the original tile position but not in the new position are evicted. The evicted region's corners are $(0,0), (0,Y_t), (X_t+d-1,0)$ and 
$(X_t+d-1,Y_t)$. 
At least one evicted element is guaranteed to have future reuse; For example, input $(0,Y_t)$ (black cell in \figref{necessary}(a)) is needed to compute the output cell $(0,Y_{t+1})$ in the future.
Eliminating the evicted input element's future reuse  by using the element  to partially compute the output cell would require holding the output cell for future computation completion, shifting the problem from the input to the output. With such partial-compute strategy, the tile must include the appropriate partial output cells. Any tile movement would similarly evict a region of the partial output needed later.

\noindent {\bf Removing our assumptions: }
(1) Because the convolutional sliding occurs only in the $X$ and $Y$ dimensions and not in the $Z$ dimension of channels, the above argument holds for multiple channels (the tile is a cuboid). 
(2) Including $(0,0)$ in the initial tile position ensures full reuse of at least $(0,0) \ldots (X_t+d-1,0)$  along the $X$-axis tile boundary. Without this assumption, there are even more elements with future reuse at the top and left of the tile. 
(3) To generalize the tile shape, we observe that because  the filters are  rectangles or squares in the $X-Y$ dimensions (e.g., $3x3$),  the {\em output} element $(0,0)$ requires the input tile to include the  three corner elements --  $(0,0)$, $(0, Y_t)$ and $(0,X_t)$ with $X_{max} > X_t \ge 0$ and $Y_{max} >  Y_t \ge 0$ --  regardless of the tile shape. Now, the above proof (for rectangular tiles) holds by ignoring the fourth corner $(X_t, Y_t)$; the same element $(0, Y_t)$ that has future reuse remains evicted, leading to sub-optimal memory traffic.

Layer Fusion~\cite{fusion} uses  tiles from another work~\cite{cong}, which are  not the full row-plane (or column-plane)  and hence do not satisfy our necessary condition. That work casts determining the tile dimensions  as an integer linear programming problem whereas we have shown that the full row-plane is necessary for full reuse. 
Unlike convolution, {\em matrix multiply}, a canonical candidate for tiling, does not have two-dimensional reuse. 
Assuming one matrix is held on-chip, any tile shape for the other
works.  In realistic scenarios, neither matrix can
fit on-chip requiring tiling of both matrices unlike our
 problem. 

Extending our necessary condition to the case of filters that are not fully-filled rectangular (e.g., dilated filters~\cite{dilated} which are checkered with holes) remains a work-in-progress. For example, with a combination of dilated filters and strides, it is possible that there are holes in the input that are never used. Naive use of rectangular tiles may result in fetching the data corresponding to the "holes" unnecessarily, thus being non-optimal. However, in such cases, it may be possible to preprocess the input to eliminate 
the holes (which are unused parts of the input), in which case the necessary condition may hold in the pre-processed input without holes.

\begin{comment}
\begin{figure}[t]
\footnotesize 
\begin{center}
\begin{tabular}{ccc}
\includegraphics[width=0.3\linewidth]{figures/necessary} & &
\includegraphics[clip, width=0.6\linewidth]{figures/sufficient}\\
(a) Necessary  & & (b) Sufficient
\end{tabular}
\end{center}
\vspace{-0.2in}
\caption{\label{fig:necessary}Necessary condition and sufficient condition}
\vspace{-0.1in}
\end{figure}
\end{comment}

\figput{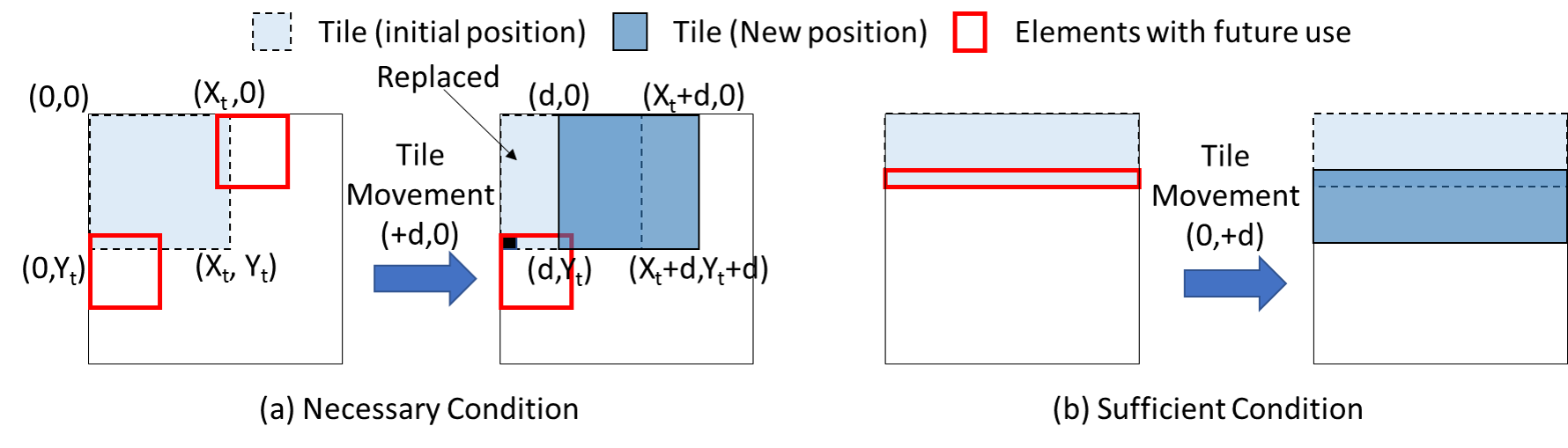}{}{Necessary condition  and sufficient condition}

\putsubsec{sufficient}{\mithuna{Sufficient condition}}  

Holding only one input map row-plane (or column-plane) is not sufficient for full reuse. Observe that a layer's output map cell is computed using many input map cells -- the output map cell's {\em dependence parents}, and that many output map cells share the same dependence parents. Fully capturing the reuse of these common parents in a single layer  requires holding on-chip all of them until all the dependent output map cells of the layer are computed.  Combined with the  above necessary condition, this  {\em single-layer sufficient condition} amounts to holding all the input map row-planes that are the  {\em dependence parents} of an output map row-plane (i.e., $k$ input map row-planes are held  if the filters dimensions are {\em k x k x m}. As mentioned in~\secref{input-reuse}, a layer in a residual CNN gets an additional input map from a previous layer. In such CNNs, an output map row-plane's dependence parents also include that layer's relevant output map row-planes. 

Assuming  each layer reads the input map from and writes the output map to off chip, the on-chip memory needs to hold all the filters and the first $k$ input map row-planes to compute the first output map row-plane (all the $n$ output map channels). 
To produce the next output map row-plane, the next set of new input map row-planes, as defined by the convolution stride, replace  those input map row-planes that are no longer needed (\figref{necessary}(b)). This strategy, employed in Eyeriss~\cite{eyeriss},  captures $k* k* n$ times input reuse requiring $2*l$ input map refetches between layers. 
To capture this reuse in a  layer, the on-chip memory must fit all the filters and $k$ row-planes of the input map (the output map is written off chip). The amount of on-chip memory needed is the maximum across all the layers.

One way to avoid the above assumption of off-chip traffic between consecutive layers is to hold on-chip, at a time, only one layer's input map and output map.
Because a layer's output map is needed only by the next layer, this strategy guarantees full reuse. 
However,  there  is often no room on-chip for the  layer's filters,
losing the
massive cross-image reuse of filters  captured by Occam.

\putsubsec{dependence-closure}{\mithuna{Dependence Closure}}  

To avoid all traffic between layers (i.e., full reuse), the single-layer sufficient condition has to be extended to all the layers' output maps.
The output maps' {\em dependence ancestors} transitively extend to include the corresponding input maps of the earlier  layers, as observed by Layer Fusion~\cite{fusion} (the solid boxes in \figref{depclosure}).  The set of dependence ancestors expands at each earlier layer in  an arithmetic sequence induced by the convolution's stride  (e.g., one row-plane of output map depends on three row-planes of input map which, in turn, {\em together} depend on five row-planes of the previous layer's input map, and  so on). The full set of the ancestors of a final output map row-plane or column-plane through all the layers to the initial input, called {\em the dependence closure},  satisfies the {\em all-layer sufficient condition}.
In residual CNNs, the dependence closure does not change due to the residual input maps  which are also fed as non-residual input maps to a previous layer and thus are present already in the closure.
As noted in~\secref{intro}, while conventional tiling typically holds only the dependence parents, this cross-layer tiling holds all the dependence ancestors.

Thus, the dependence closure  defines the shape and size of the optimum tile to achieve full reuse.
In contrast, Layer Fusion holds the ancestors of an output map tile that does not satisfy the necessary condition. As such, Layer Fusion captures between $k * n+2$ and $k*k*n$  times input map reuse with the same  on-chip memory as Occam (i.e., between $2*l$ and $k*l$ 
input map refetches for $l$ layers assuming all the layers' input maps and output maps are the same size). As noted in~\secref{intro}, while conventional tiling typically holds only the dependence parents, this cross-layer tiling holds all the dependence ancestors.

\figput[0.7\linewidth]{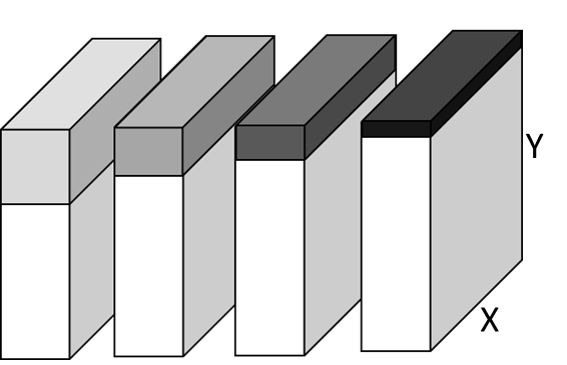}{}{Dependence Closure}

\figputW{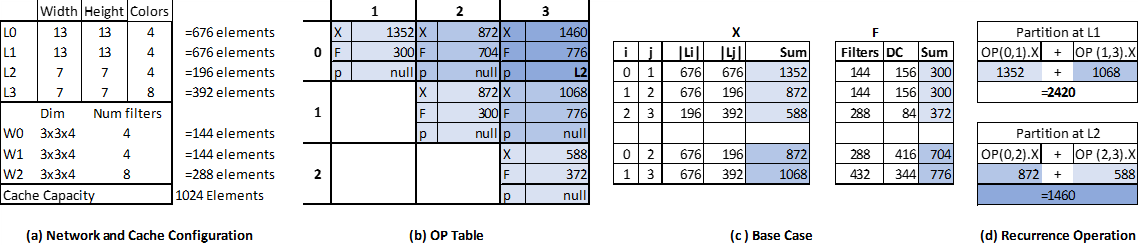}{}{Walkthrough Example of Optimal CNN Partitioning}

Under dependence closure, execution proceeds by computing only the required number of output map row-planes in each layer as per the above arithmetic sequence to produce the first row-plane of the final output. While the initial input is read from off-chip and the final output is written off-chip, 
the full dependence closure data (i.e., the relevant partial output map of all the layers) is held  on-chip. 
Going from the final output's top row-plane to the next row-plane, the dependence closure also slides down at each layer by the layer's convolution stride (see the broken boxes in \figref{depclosure}). Because the stride is much smaller than the arithmetic sequence values, the dependence closures of consecutive final output row-planes overlap greatly. Occam captures this overlap 
 by holding on-chip the entire closure  until needed and replacing only the unneeded, older parts of the  closure  with newer  parts. 
Because Layer Fusion does not satisfy our necessary condition, it does not hold the entire closure. Instead, Layer Fusion proposes
to recompute the missing parts of the closure (and to refetch the relevant input). 
However,  the large overlap implies significant amount of recomputation added  to the already compute-intensive CNNs.

To produce the final output's second row-plane, the first layer discards  (1) the top input map row-planes  no longer needed, replacing them with the next set of  input map row-planes as per the convolution stride, and (2) the output map row-planes in the temporary space that are no longer needed, overwriting them with the next set of output map row-planes. Thus, the temporary space acts conceptually as a  circular buffer for the relevant input map and output map row-planes, similar to Layer Fusion. This circular buffer repeatedly reuses its space to hold the dependence closure of subsequent final output row-planes, conserving on-chip memory capacity. 
Later layers follow the same strategy (i.e., each layer has its own circular buffer of size defined by the arithmetic sequence). Because the computation crosses all the layers to produce each final output row-plane, all the layers' filters need to be held on-chip along with  the dependence closure.

The local operations for batch normalization, pooling, bias addition, and  ReLU (\secref{input-reuse}) occur as
part of each layer's computation without affecting the dependence closure. 

\putsubsec{partition}{\mithuna{Optimal Partition}}

Real CNNs' full set of filters and dependence closures are too large to fit in one chip's on-chip memory, or cache. Therefore, we partition a given CNN into sets of contiguous layers where each partition's dependence closure fits in the cache. Each partition reads its input once from and writes its output to off-chip memory. Layer Fusion~\cite{fusion} explores such partitioning  via brute-force search which is infeasible for modern
CNNs which may have more than 100 layers. Instead, we propose a dynamic programming (DP) algorithm to guarantee the least off-chip traffic for a given CNN and cache capacity while ensuring that filters
remain   cache-resident to capture cross-image filter reuse. 
We use a running example (\figref{walkthrough}) for illustration.

Our formulation uses the following definitions:

\noindent
$\bullet$ 
$L_i$ and $W_i$ define the input feature map data and filters, respectively, for the $i^{th}$ layer. $|L_i|$ and $|W_i|$ define the number of elements in $L_i$ and $W_i$, respectively, independent of data format (e.g., FP32, FP16, INT8). $L_0$ is the input image. \figref{walkthrough}(a) shows the four feature maps ($L_0$ through $L_3$) and  the three layers' weights ($W_0$ through $W_2$) for our example. 

\noindent
$\bullet$ 
We assume the  cache can hold $C$ elements ($C=1024$  in 
\figref{walkthrough}(a)).

\noindent
$\bullet$ DC(i,j) defines the dependence closure of one row-plane of the output feature map in $L_j$ extending back to the feature map of layer $L_i$ where $(0\le i < j \le n)$ and $n+1$ is the number of layers. Thus, 
the end-to-end dependence closure defined earlier in \secref{dependence-closure} has $|DC(0,n)|$ 
elements. 

\noindent
$\bullet$  We define a $SPAN(i,j)$ as the convolution computations starting with $L_i$ as the input and ending with $L_j$ as the output.

For on-chip resident filters, the minimum total footprint of {\em all} convolutional layers
is the sum of all weights ($\Sigma_{i=1}^{n-1}|W_i|$) and the full dependence closure
($|DC(0,n)|$). If this footprint fits in the cache, then optimal
operation needs no partitioning. 
Otherwise, the layers must be partitioned into {\em spans} such that
(1) each $SPAN(i,j)$ satisfies the capacity constraint that the dependence closure  (i.e., $|DC(i,j)|$) and  weights (i.e., $\Sigma_{k=i}^{j-1} |W_k|$ ) of the span must fit in the cache, and (2) the total amount of data transferred off-chip at the partition boundaries is minimized. 
\begin{comment} IPDPS
Even in the degenerate case of a single-layer span, Occam offers modest, albeit reduced, bandwidth improvements.
We later address the final case, where even a single layer does not fit in the cache. 
\end{comment}

\begin{mydef}
For a CNN with  $n$ convolutional layers, we define a partition boundary set (PBS)
as a set $P = \{p_1,p_2,\ldots,p_{k-1}\}$ that specifies a partitioning of the  CNN into $k$ spans --- SPAN($0,p_1$), SPAN($p_1,p_2$), $\ldots$,
SPAN($P_{k-1},n$). Here, we assume that the elements of $P$ are strictly increasing integers (i.e., if $i<j$ then $p_i < p_j$) that  lie between 0 and $n$.
For uniform naming of partition boundaries, we define $p_0=0$ and $p_k = n$.
\end{mydef}

In a valid PBS, each  span's footprint fits in the cache:  
\begin{align}
\forall 0 \le i < n, (|DC(p_i, p_{i+1})| + \Sigma^{p_{i+1}-1}_{k=p_i}|W_k|) < C \eql{0}
\end{align}
Further, for any SPAN($p_i,p_{i+1}$) specified by such a PBS, the associated computation may run on a single chip. 
The only off-chip transfers are 
(1) reads of the input layer which transfers
$|L_{p_i}|$ elements from off-chip memory (or upstream chips), and (2) writes of the output layer which transfers 
$|L_{p_{i+1}}|$ to off-chip memory (or downstream chips). 

Informally, our problem fits DP because optimally partitioning a network into two would require that the ``left'' and ``right'' sub-partitions themselves be optimal.
This claim is sound because of the {\em optimal substructure} property of the problem~\cite{cormen}, which can be proved easily by contradiction as follows. If the optimal solution to the $SPAN(i,j)$ uses suboptimal solutions (i.e., partitions that yield more transfers) to a recursive subproblem, using the optimal solution for the subproblem yields fewer overall transfers than the optimum.

In our DP table $OP[n,n]$, each cell $OP[i,j]$ holds three fields of information on the optimal partition of an arbitrary span from $L_i$ to $L_j$: 
(1) $p$, the feature map that marks the optimal point of partition between $L_i$ and $L_j$, and (2) $X$, the optimal number of transfers. 
\figref{walkthrough}(b) shows the OP table for our example, wherer $F$ denotes  the footprint of the largest sub-span within the span as per the optimal partition.

\noindent
{\em DP base case:} 
In the base case, if the entire footprint (filters + DC) of a $SPAN(i,j)$ fits in cache (Eqn.~\eqref{eq:0}), we initialize  the 
number of transfers as the bare minimum --- every element in the input and output layers
$i$ and $j$, respectively, is read and written exactly once (Eqn.~\eqref{eq:1}). \figref{walkthrough} shows the example's footprints (F), even though they are not tracked in the DP table.
Because the filter transfers are amortized to zero over multiple images,
the filters are not counted in the transfers. 
Further, because no partitioning is necessary, $p$ is set to $null$ (Eqn.~\eqref{eq:3}). 
\begin{align}
    OP[i,j].X &= |L_i| + |L_j| \eql{1}\\
    OP[i,j].p &= null\eql{3}
\end{align}

If single-layer spans ($SPAN(i, i+1)$) fit in the cache,
we initialize  $OP[i,i+1]$, $0\le i<n$, using the above assignments, as shown with light shading in \figref{walkthrough}(b) and \figref{walkthrough}(c).

\begin{comment} IPDPS
If even a single layer does not fit, then chip-residence for filters is not possible and  weights must be included in the transfers (Eqn~\eqref{eq:9}). In this special case, the transfers are a lower bound; depending on the implementation (i.e., its access patterns), either the input or filters may be transferred more than once. 
And, the layer remains a partition by itself. 
\begin{align}
  OP[i,i+1].X  &= |L_i| + |L_{i+1}| + |W_i| \eql{9}
\end{align}
\end{comment}

\noindent 
{\em Recurrence for other cases:}
Our DP algorithm solves the optimal partition problem in a bottom-up fashion by increasing the span length beyond the base case of 1. 
A longer span either fits in the  cache or not.
The first case is treated exactly as the base case. OP[0,2] and OP[1,3] illustrate this case in \figref{walkthrough}(b) and \figref{walkthrough}(c)).
For the second case, we  consider every possible partition of $SPAN(i,j)$ and pick the  partition point $p$  that yields the fewest transfers in the two resulting sub-spans, $SPAN(i,p)$ and $SPAN(p,j)$. 
\begin{align*}
p_{opt} &= p \mid \forall (i < k < j) \\
&(OP[i,p].X + OP[p,j].X) \le (OP[i,k].X + OP[k,j].X)
\end{align*}

The choice of the optimal partition point is shown in \figref{walkthrough}(d). The algorithm compares the two choices ($L_1$ and $L_2$) and chooses $L_2$ which results in fewer transfers.

We define the solution to longer spans $SPAN(i,j)$ as:
\begin{align}
	OP[i,j].X &= OP[i,p_{opt}].X + OP[p_{opt},j].X \eql{4}\\
    OP[i,j].p &= p_{opt}\eql{6}
\end{align}

The above recurrence accurately tracks the number of the off-chip transfers (i.e., $X$) of each of the two resulting spans (Eqn.~\eqref{eq:4}).
Further, saving the partition points ($p$)  facilitates reconstruction of the 
final optimal PBS (Eqn.~\eqref{eq:6}).
\figref{walkthrough}(b) shows the full computation of OP[0,3] which yields the optimal partitions: SPAN(0,2) and SPAN(2,3).

Finally, a divide-and-conquer algorithm, instead of DP, would not be  efficient here because of the numerous overlapping sub-problems. 
% -- a well-known symptom that suggests DP. 
For example, the subproblem $OP(3,6)$ would  be revisited when  examining larger spans (e.g., $OP(2,8)$ and  $OP(3,7)$). DP avoids
such recomputation by memoizing the solutions in the $OP$ table. 

\noindent 
{\bf Extensions:} The above algorithm extends easily to handle (1) residual connections
such as those used in ResNets~\cite{resnet}, and (2) batched computation for inference on a minibatch of multiple images. 
% Due to lack of space, we  do not describe the details. 

Residual connections
effectively read and aggregate values from upstream layers which results in additional transfers. 
Residual connections interact with Occam in one of two ways:  either the residual connection does not span any partition boundary or it does.
In the first case, 
Occam guarantees that the residual reads impose no additional
off-chip transfers because the residual  values 
are already in the dependence closure.
In the second case, 
the residual values result in additional transfers as the values  must be written out to and read back from memory. 
These additional transfers require a minor change
to Eqn.~\eqref{eq:4}  as follows.
$$	OP[i,j].X = OP[i,p_{opt}].X + OP[p_{opt},j].X + 2\times|L_{source}|$$

Batched inference is  used commonly in  CNNs (say with a minibatch of $b$ images). 
The only change  is that feature map transfers and footprint (\eqref{eq:0})  increase proportionally to $b$ whereas filter transfers 
and footprint remain unchanged because the entire minibatch uses the same chip-resident filters. 
Accordingly, Eqn.~\eqref{eq:1} is modified to Eqn.~\eqref{eq:7}.
\begin{align}
    OP[i,j].X &= {b\times(} |L_i| + |L_j| {\bf)} \eql{7}
\end{align}

\noindent 
{\bf Complexity:}
The DP algorithm is used to optimize partitions offline (like compiler optimizations). 

The algorithm is of asymptotic complexity $O(n^3)$ as there is potentially $O(n)$ work to find the optimal partition point for 
up to $O(n^2)$ spans in the $OP$ table. In practice, the algorithm's 
runtime is less than a second on a laptop, even for the largest network we consider (Resnet-152). Given that an optimal  solution can be computed efficiently, exploring heuristics for the  problem is probably not warranted.

\figput[0.8\linewidth]{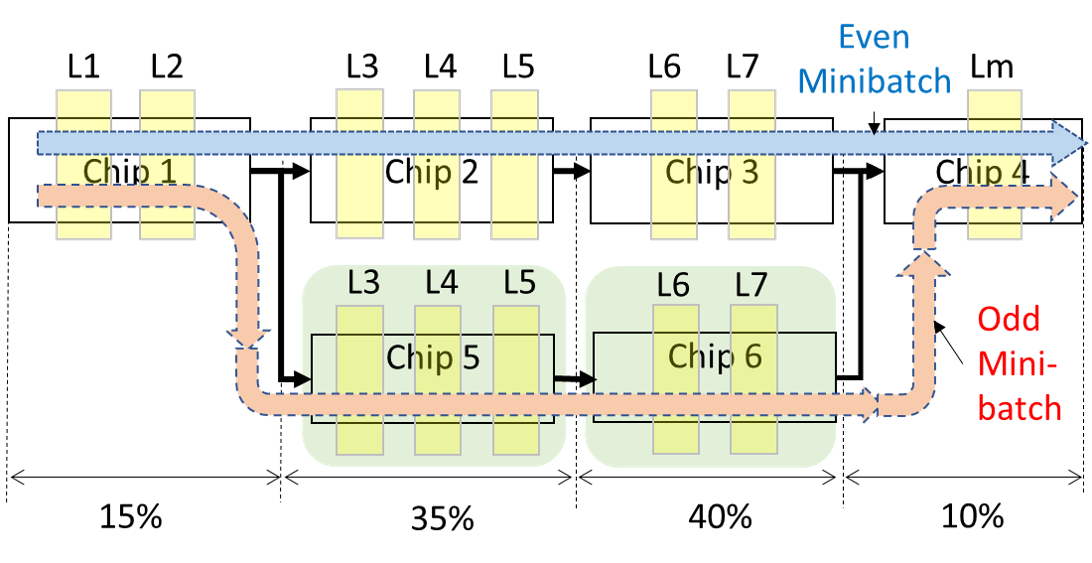}{}{Pipeline for inter-image filter reuse}
 
\putsubsec{smbp}{Staggered Asynchronous Pipelining (STAP)}  
Occam captures the enormous inter-image   filter reuse by placing each partition -- its filters and dependence closure  --  in a separate chip (e.g., a GPU, TPU, or  FPGA). 
The partitions form a multi-chip {\em asynchronous} pipeline both to capture full inter-image  reuse and to achieve high throughput (see \figref{pipeline}).

The input images pass through the pipeline requiring off-chip access only
for each partition's (pipeline stage's) input and output feature maps but not the filters (i.e., full cross-image reuse). In contrast, the input-stationary approach~\cite{eyeriss} refetches the filters for every input image (e.g., TPU).  

In each stage, the earlier output map cells are written (or communicated) while the later cells are produced, so that  data transfers are hidden  under the producer stage's heavy computation except for a small {\em initial} part (\secref{perf}). While Occam's pipeline achieves optimal transfers,
other techniques~\cite{brainwave,Kung} employ ad hoc partitions. 

\mithuna{Because Occam's partitions may result in an unbalanced pipeline, the throughput is limited by the bottleneck stage. However,
the  latency is unaffected due to {\em asynchronous} pipelining  if the job arrival rate is under the bottleneck  rate (asynchronous  stages do not wait for clock edges as do synchronous stages). 
For example, in an unbalanced, 4-stage Occam pipeline with 15-35-40-10 latency-units per stage, the latency is 
100 units and the throughput is  1/40 $units^{-1}$.}

The throughput can be increased by replicating the bottleneck  stages. In our example,  
stages 2 and 3 are each replicated for a throughput of one inference per 20 units (\figref{pipeline}). 
These stages use the $i^{th}$ replica for the $i^{th}$  input mini-batch (i.e., parallelize across input mini-batches in a staggered manner and {\em not}  within each mini-batch).
Importantly, STAP achieves better throughput {\em without} changing Occam's optimal tiles.

We consider other forms of parallelism, typically used in training. Model parallelism splits  one or more~\cite{weird-trick,layer-wise,hypar} layers into multiple chips and must (1) copy the input to each chip and  (2) merge  the chips' results for the next layer. These overheads are worthwhile to reduce filter update traffic in  training, but not in inference where filters are not updated and data parallelism achieves the same effect without the overheads.
Both Layer-wise parallelism~\cite{layer-wise} and HyPar~\cite{hypar} choose between model and data parallelisms for each layer, but model parallelism is not relevant for inference.
Data parallelism is {\em orthogonal} to STAP which achieves higher {\em throughput}. For  lower {\em latency} per image, each mini-batch can exploit data parallelism by replicating the {\em entire} pipeline
without affecting Occam's transfer optimality.
(e.g., \figref{pipeline}). Each replica would work on half the mini-batch. Thus, data parallelism does not change Occam's off-chip transfer optimality. 

Finally, GPipe~\cite{gpipe} exploits pipelined parallelism for training. Gpipe uses heuristic-based partitioning (i.e., not optimal, like Occam) and targets minimizing the training pipeline loop from forward pass to back propagation -- a problem that is not relevant to Occam.

\putsec{method}{Methodology}
We use two implementations -- a software-based and an FPGA-based --  to evaluate Occam.

\noindent
{\bf Software implementation: }
We implement Occam in CUDA which is used prevalently to implement CNNs. 
Because Occam requires changes to the CUDA kernels of CNN implementations whereas commercial CUDNN frameworks are not open source, we choose   {\em Convnet}~\cite{alexnet}, an open-source framework.
Like the latest CNNs, we use 8-bit integers (INT8) which both reduces the data volume and enhances the compute parallelism compared to 32- or 16-bit floating point data.

\begin{comment} IPDPS
While the on-chip memory could be a software or hardware cache, the latter may incur conflict misses despite high associativity whereas our partitioning considers only the cache capacity. We layout each partition's filters and  dependence closure (\secref{dependence-closure}) sequentially to fit in the chip's cache without capacity or conflict misses.
\end{comment}

We implement our DP algorithm as a standalone JavaScript  application that takes as input the network parameters 
and produces the optimal partitions and tile dimensions. We then feed these outputs to Convnet.

\noindent
{\bf Choice of Simulation: }
While our CUDA implementation can run on a real GPU,
Nvidia GPUs exhibit undocumented cache 
behavior that makes it hard to exploit reuse. 
For example, our tests revealed that both prefetch (up to 256 KB but not more) and early eviction (e.g., data read once but not written  is evicted before the second touch within the same kernel and even when the footprint fits in the cache) make capturing reuse all but impossible (also observed in other studies~\cite{microbench1}). More importantly, the hardware offers no mechanisms to disable these behaviors when they hurt performance. Therefore, we instead 
use a simulator (GPGPU-sim) where the cache behaves as expected. Our FPGA implementation  does not have these problems (\secref{fpga-results}). 

\noindent
{\bf Simulated system:}
{\em Our goal is to show that Occam improves performance over the best current system.} 
Such a system uses ``minibatches" to perform batched inference on an accelerator  like Google's TPU or the latest GPUs.
\mithuna {We consider an Nvidia Volta-like  accelerator with 140K multiply-accumulate units, 18-MB on-chip cache, and 800-GB/s memory bandwidth processing 32-image minibatches.

We consider such an  accelerator with 32K multiply-accumulate units, 24-MB on-chip cache, and 180-GB/s memory bandwidth processing 32-image minibatches.
These  resources are shared across  a minibatch of 32 images is shown in \figref{occamaccel}. 
However,  simulation of entire minibatches would be impractically slow. 

For practical simulation times, we carefully scale the simulated GPU to match a slice of the above system that performs a single inference out of the 32-image minibatch. This scaling reduces the compute bandwidth by a factor of 32 because the work is proportional to the number of inferences. Further, because the simulated GPU incurs instruction overheads that an accelerator does not, the compute bandwidth is increased to 15K multiply-accumulate units. (For example, ResNet requires around 200 multiply-accumulate operations per memory byte. But a GPU requires around 3000 instructions per byte.)

For cache sizes and memory bandwidth, 
the scaling has to ensure that (1) the partition's filters are chip-resident and (2) the filters are shared, in cache capacity and memory bandwidth, across all 32 images in the minibatch. Accordingly, 
we scale the cache to hold one image's feature map data and the full filter data. Because this number varies for each partition, we use a calibration based on AlexNet which yields a cache size of 3 MB. We validated our scaling methodology by simulating 1, 2,  and 4 slices 
with appropriate scaling and verifying performance within 3\% of one another.} We use a similar calibration for memory bandwidth reducing it by a factor of 6 (to around 133 GB/s). 
Finally, we vary the cache size to show that Occam works for other cache sizes. 
The simulated GPU's parameters are shown in~\tabref{gpu_config}.

\begin{table}[t]
{\small
\begin{center}
\begin{tabular}{|l|l|}
\hline
\# Streaming Multiprocessors                                              & 248   \\ \hline
Pipeline Width                                                            & 128 (INT8)     \\ \hline
Number of Registers / SM                                                  & 65536   \\ \hline
Shared Memory / SM                                                        & 98 KB   \\ \hline
L2  Cache (shared)                                                        & \begin{tabular}[c]{@{}l@{}}3 MB, 128B line,\\ 12 Banked, 16 Way Assoc.\end{tabular} \\ \hline
Memory Bandwidth   & 133 GB/s   \\ \hline
\end{tabular}
\end{center}
}
\caption{Scaled hardware parameters for a single image}
\label{tab:gpu_config}
\vspace{-0.4in}
\end{table}

While our on-chip residence requires multiple chips (STAP in~\secref{smbp}), 
we simulate a single GPU to simplify our implementation. 
To emulate chip-resident filters in our one-GPU simulations, we pre-touch the filters for each partition (in Occam or Layer Fusion) before the partition executes.

\noindent
{\bf Schemes: }
We implement three schemes; a layer-by-layer base case,  Occam, and Layer Fusion. \mithuna{The filters of any single layer fits in the cache in the base case, which, like Eyeriss~\cite{eyeriss}, captures {\em all} reuses, except cross-layer, {\em even without tiling}.} We compare to Layer Fusion which includes cross-layer reuse (due to partitioning {\em and} tiling). As such, Layer Fusion subsumes other techniques that have partitioning alone (e.g., Brainwave~\cite{brainwave}).
Occam's chip-resident approach uses multiple chips for a given network. 
To ensure an equal-cost comparison, our base case uses the same number of chips as Occam for throughput via replication. 
Layer Fusion's implementation is similar to Occam's, except with different tile shapes and sizes. 
Layer Fusion's exhaustive search for partitions is infeasible for large networks (e.g., $2^{34}$ choices in  {\em ResNet-34}). As such, we use Occam's optimal partitions for Layer Fusion and choose the largest square tile whose dependence closure for a given partition would fit in the cache (a different tile size for each partition). Because the tiles are  sub-optimal even though the partitions are optimal, Layer Fusion does not capture full reuse. Recall from~\secref{dependence-closure} that Layer Fusion employs recomputation for the evicted parts of the dependence closure. By using Occam's partitions, Layer Fusion also uses the same number of chips.  
We  verified the functional correctness of these schemes by comparing against unmodified Convnet.

\figput[0.8\linewidth]{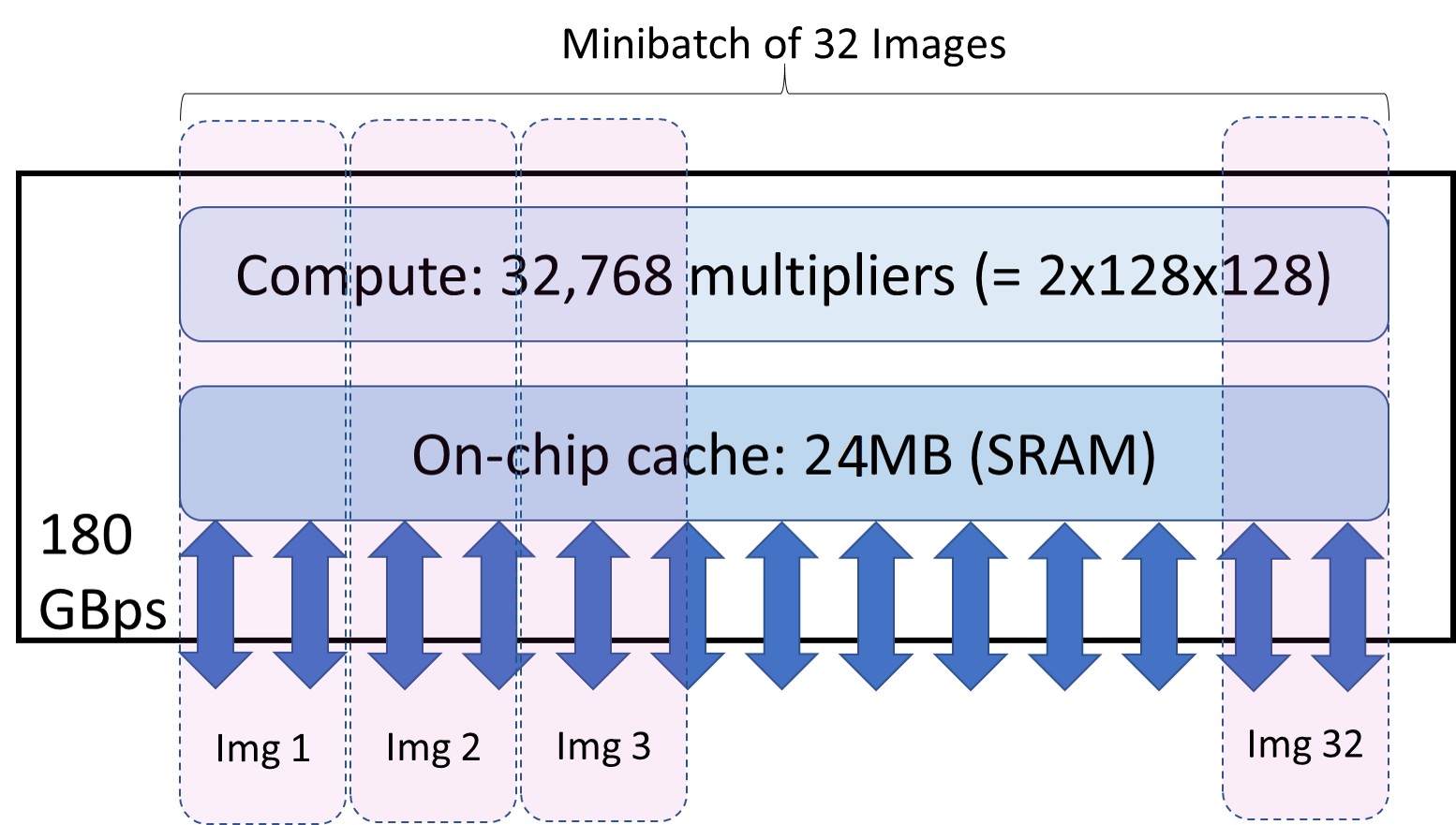}{}{Target hardware and software setup. Simulation covers a single inference from a minibatch}

\tabputW{benchmarks}{
\small
\begin{stripetabular}{|c|c|p{1.0in}|p{2in}|p{2in}|}
\hline
Network&Layers &Partitions & \multicolumn{2}{|c|}{ Partition boundaries and Tile sizes ($P_{start}, P_{end}, TileDim$)}  \\
 & &   & Occam ($TileDim \times Row width$)& Layer Fusion ($TileDim \times TileDim$) \\
\hline

AlexNet & 8 &  0,8   &(0,8,2) &(0,8,3) \\
VGGNet & 19 & 0,6,11,12,14,15,16,18 & (0,6,6)(6,11,1)(12,14,6)(16,18,7) & (0,6,20)(6,11,5)(12,14,8)(16,18,7) \\
ZFNet  & 8  &  0,5,8 & (0,5,13)(5,8,6) & (0,5,13)(5,8,6)\\
ResNet-18  & 18 & 0,12,15,16,17,18 & (0,12,2)(12,15,7)  & (0,12,4)(12,15,7)\\
ResNet-34  & 34 & 0,15,21,26,29,30,31, 32,33,34 &(0,15,3)(15,21,2)(21,26,2)(26,29,7) & (0,15,7)(15,21,5)(21,26,5)(26,29,7)\\
ResNet-50  & 50 & 0,21,30,37,42,43,45, 46,48,49,50 & (0,21,2)(21,30,2)(30,37,2)(37,42,7) (43,45,7)(46,48,7)&(0,21,9)(21,30,5)(30,37,5)(37,42,7) (43,45,7)(46,48,7) \\
Resnet-101 & 101 &  0,21,30,37,45,52,60,67, 75,82,90,93,94,96,97, 99,100,101& (0,21,1)(21,30,2)(30,37,2)(37,45,2) (45,52,2)(52,60,2)(60,67,2)(67,75,2) (75,82,2)(82,90,2)(90,93,7)(94,96,7) (97,99,7)&(0,21,9)(21,30,5)(30,37,5)(37,45,5) (45,52,5)(52,60,5)(60,67,5)(67,75,5) (75,82,5)(82,90,5)(90,93,7)(94,96,7) (97,99,7) \\
Resnet-152 & 152 & 0,21,36,43,51,58,66,73, 81,88,96,103,111,118, 126,133,141,144,145,  147,148,150,151,152 & (0,21,1)(21,36,4)(36,43,2)(43,51,2) (51,58,2)(58,66,3)(66,73,3)(73,81,2) (81,88,2)(88,96,2)(96,103,2) (103,111,2)(111,118,2)(118,126,2) (126,133,2)(133,141,2)(141,144,7) (145,147,7)(148,150,7)&(0,21,9)(21,36,8)(36,43,5)(43,51,5) (51,58,5)(58,66,5)(66,73,5)(73,81,5) (81,88,5)(88,96,5)(96,103,5) (103,111,5) (111,118,5)(118,126,5) (126,133,5)(133,141,5)(141,144,7) (145,147,7) (148,150,7) \\
\hline
\end{stripetabular}
}{Benchmark Characteristics}

\noindent {\bf Benchmarks:}
\tabref{benchmarks} shows the networks used as benchmarks. We repeat each run thrice to capture any statistical variations (which are little to none). 
\tabref{benchmarks} also lists the partitions (same for Layer Fusion and Occam) and the tile sizes for each scheme. We run some training runs to generate each network's filters. Because we are not interested in accuracy (studied elsewhere), we do light training to save time. 
We use the input images provided with Convnet for all our runs. {\em We simulate full network execution except the  fully-connected layers.}

\putsec{results}{Results}
We present three sets of results: analytical, simulation, and FPGA. 
The analytical results show  Occam's optimal tile and partitions for the benchmark networks and the traffic savings as tracked by our algorithm (\secref{partition}). The simulation results compare the execution times and energy for Occam and Layer Fusion. The FPGA results compare performance for Occam and the base case.  

\tabput{misses}{
\small
\begin{tabular}{|l|rr|rrr|rr|}
\hline
Network & \multicolumn{2}{|c|}{Miss (Meas.)} & \multicolumn{3}{|c|}{Miss (Model-predict)} & \multicolumn{2}{|c|}{Norm. Insts.} \\
& \multicolumn{2}{|c|}{(Base = 1)} & & &  & \multicolumn{2}{|c|}{(Base = 1)} \\
& Occ. & L.Fu. & Base & Occ. & L.Fu. & Occ. & L.Fu.\\
\hline
Alex & 0.05 & 0.10 & 1.0 & 0.05 & 0.10 & 1.05 & 1.70 \\
VGG & 0.16 & 0.10 & 0.6 & 0.06 & 0.06 & 1.03 & 1.29 \\
Zf & 0.07 & 0.07 & 1.0 & 0.06 & 0.06 & 1.05 & 1.05 \\
Res-18 & 0.03 & 0.06 & 1.0 & 0.04 & 0.04 & 1.05 & 1.72 \\
Res-34 & 0.04 & 0.06 & 1.0 & 0.04 & 0.04 & 1.04 & 1.77 \\
Res-50 & 0.03 & 0.05 & 1.0 & 0.04 & 0.05 & 1.03 & 1.36 \\
Res-101 & 0.03 & 0.04 & 1.0 & 0.03 & 0.04 & 1.04 & 1.40 \\
Res-152 & 0.03 & 0.04 & 1.0 & 0.03 & 0.04 & 1.04 & 1.37  \\
\hline
Mean &0.05 &0.06 &0.9 &0.04 &0.05 &1.04 &1.44\\
\hline
\end{tabular}
}{Normalized LLC Miss and Instruction Counts}

\putsubsec{netanalysis}{Analytical results}
In~\tabref{benchmarks}, we present the optimal partitions and tile dimensions for our  networks for 3-MB  on-chip memory. 
While the tile dimensions for  Occam and Layer Fusion are different, we use Occam's partitions for Layer Fusion (\secref{method}).  The partitions are shown using the start layer for each  (e.g., AlexNet uses a single partitions from layers 0 through 8). 
For each partition, we show the corresponding tile sizes for both Layer Fusion and Occam. 
The tile sizes are shown as triplets of the form $(pBegin, pEnd,$ $pTileDimension)$ where pBegin and pEnd are the layers at the beginning and the end of the partition.   Layer Fusion's tiles are square-shaped ($TileDim\times TileDim$). In Occam, the tile dimension corresponds to the number of full rows ($TileDim\times RowWidth$).
For a realistic capacity of 3 MB, Occam is able to achieve many multi-layer partitions which capture inter-layer reuse.  

Recall that each partition's filters reside on-chip for all the schemes. 
Our algorithm calculates the capacity for  each partition's filters and dependence closure. \figref{capacitysplit} shows this capacity  split for {\em ResNet152}.
For all of our networks, most of the on-chip capacity goes to the filters and a small fraction to the dependence closures.
This result highlights the importance of our sufficient condition (our second contribution).  The large capacity for the filters  saves significant off-chip traffic due to massive cross-image filter reuse (our fourth contribution).

\figput{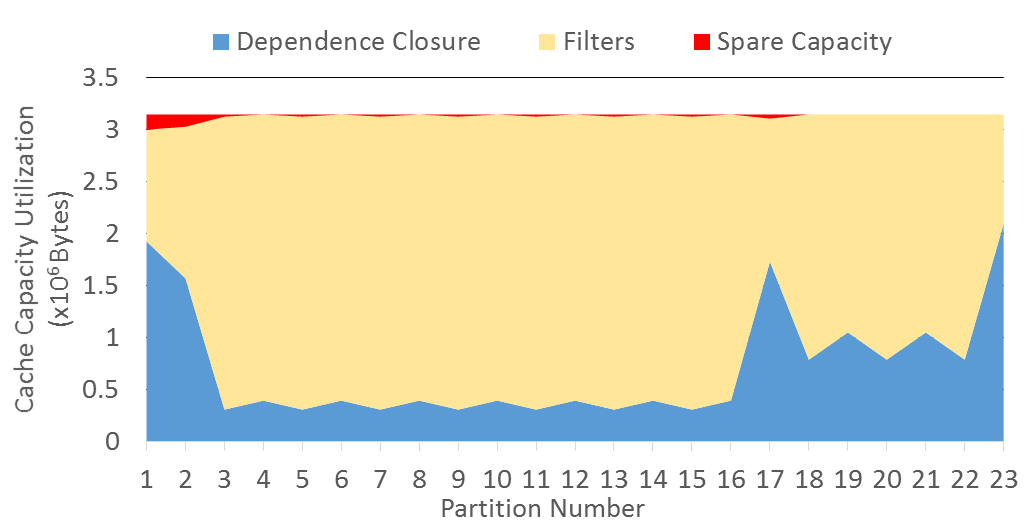}{}{Capacity Split  for {\em ResNet152}}

\putsubsec{expt}{Simulation Results}
\putsubsubsec{perf}{Performance:}
\figref{main} shows the speedup (Y-axis) for Occam and Layer Fusion over  the 
baseline for various CNNs (X-axis). 
In addition to the individual networks, \figref{main} also shows the geometric mean speedup across all networks (rightmost bars). Occam uniformly outperforms the baseline with a mean speedup of 2.06x over all networks. The speedups are higher for the larger, more recent networks such as the ResNets. 
These speedups are a direct result of the large reduction in the miss counts achieved by Occam (21x on average) while avoiding instruction bloat (measured miss counts normalized to the measured baseline's  in \tabref{misses}). This miss reduction is due to our necessary condition which leads to our specific tile shape and our dynamic programming algorithm which minimizes off-chip traffic (our first and third contributions).  
\figput{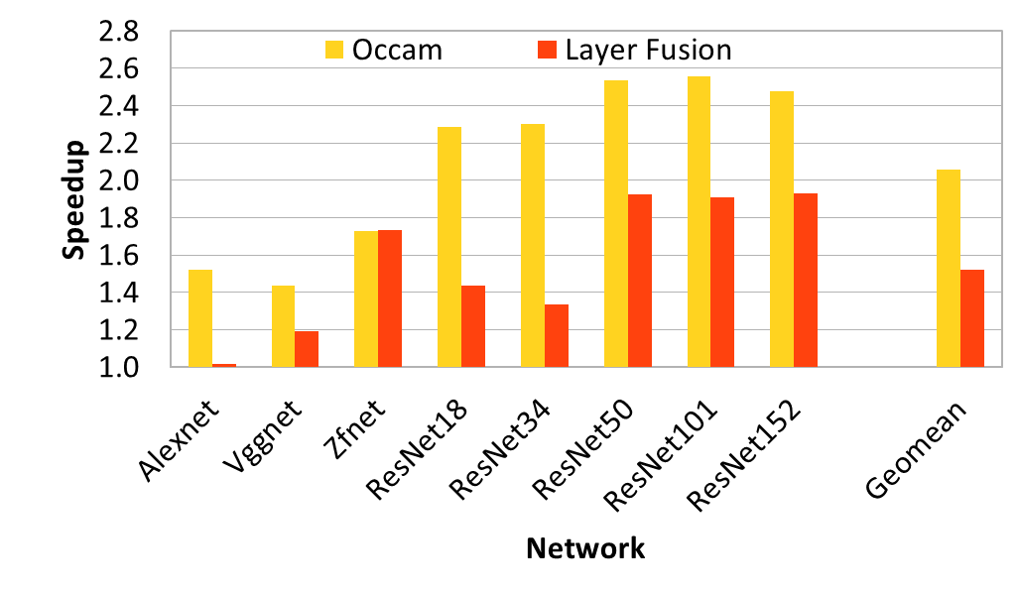}{}{Occam's Kernel Execution Speedup}

\tabref{misses} also shows that our model-predicted miss counts (normalized to the measured baseline's) closely match those seen in simulation measurements. The only exception is VGGNet where there are a few individual layers that are too large to fit in the cache. Recall that Occam uses a lower-bound estimate for such cases.

Layer Fusion also achieves speedups, 1.52x on average, due to lower traffic. 
However, its speedups are lower than Occam's due to its sub-optimal tiles which induce high recomputation cost. 
The recomputation cost is quantified in the normalized instruction count shown in \tabref{misses}. The three networks where Layer Fusion lags the most behind Occam ({\em AlexNet, ResNet18,} and {\em ResNet34})
are indeed the three networks with the highest relative instruction overhead. {\em Zfnet} is a special case where the entire output and its dependence closure fits in the cache for each partition. In this case, Occam and Layer-fusion are effectively equivalent. Despite its sub-optimal tiles, Layer Fusion seem similar to Occam in miss counts because Layer Fusion recomputes instead of incurring  extra misses.

Finally, Occam's latency penalty for the {\em initial} inter-chip transfer has minimal impact on performance (\secref{smbp}). The typical-case PCIe latency of 30$\mu$s per partition (PCIe latency varies from 10$\mu$s to 50$\mu$s~\cite{cpcie})  results in slightly reducing the average 2.06x speedup  to approximately 2.01x. (Subsequent transfers are hidden under computation due to pipelining.)

We discuss next  the energy penalty of inter-chip communication.

\putsubsubsec{energy}{Energy:}

We did not use GPUWattch~\cite{gpuwattch} due to its several quirks which lead to evidently incorrect results. For example, as part of minimizing the mean-square error with its benchmarks, GPUWattch assigns large scaling factors  without any physical rationale (e.g., memory write energy scaled down by $10^{4}$). These factors leads to obviously incorrect results when simulating CNNs (e.g., memory system energy is approximately 0.25\% of total energy). Instead, we use TPU's compute energy
of 0.43 pJ/op~\cite{TPU} and GDDR5 DRAM energy of 6pJ/bit or 48 pJ/B~\cite{micron} (roughly 100x more expensive than compute~\cite{Dally-100x}). 

\figput[\linewidth]{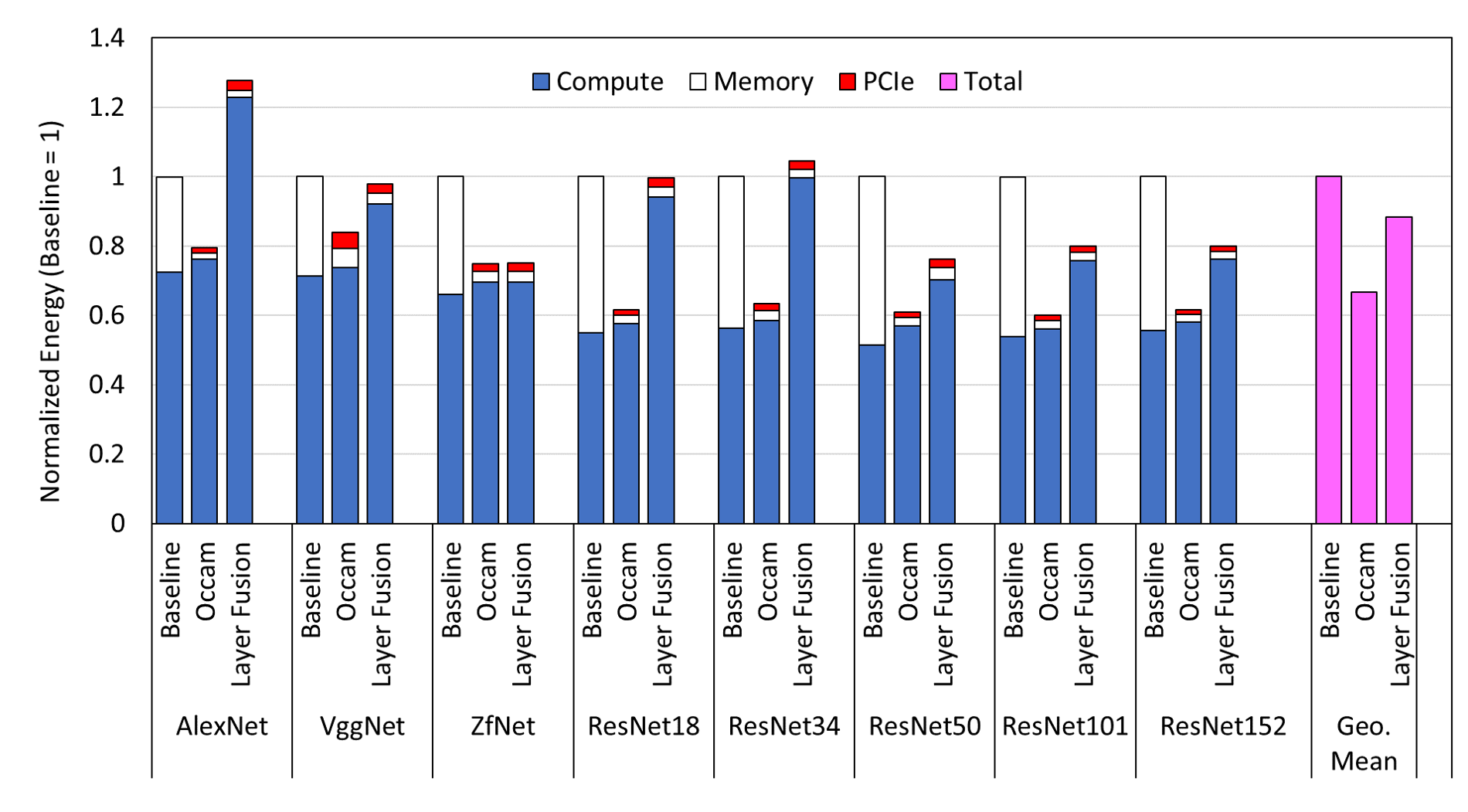}{}{Occam's kernel energy savings}

\figref{energy} shows the energy breakdown into compute, memory, and chip-to-chip PCIe components for the baseline,
Occam, and Layer  Fusion.  The PCIe component is not
applicable to the baseline which is not chip-resident and runs one network entirely on one chip; for equal cost, the baseline uses the same number of chips as Occam for parallelism (\secref{method}). 
For an equal-cost comparison, we assume chip residence even for Layer Fusion which then incurs PCIe energy like Occam (\secref{method}). 
In the baseline, compute and memory energy are split roughly on  average as 65:35.  
Occam's compute energy is slightly more than the baseline's as shown by the instruction counts (\tabref{misses}). Occam drastically cuts memory transfers (21x on average, in \tabref{misses}) but incurs the extra energy  of chip-to-chip transfers at partition boundaries.
The net effect of these factors is 33\% average reduction in energy because the energy cost/bit for DRAM and PCIe are similar (6pJ/bit ~\cite{micron,cpcie}). 
Finally,  Layer Fusion's memory energy saving due  to fewer
transfers (\tabref{misses}) are offset by its  
 significant recomputation  overhead induced by its sub-optimal tiles, resulting in a net energy saving of 12\% on average.
 
As we increase the  cache size from 3 MB to  6 MB, Occam's speedups improve (not shown). 

\begin{comment}

\putsubsubsec{csense}{Sensitivity to Cache Capacity:}
To evaluate Occam's sensitivity to cache capacity, and to confirm that Occam's techniques apply broadly, \figref{csense} presents the speedups (Y-axis) for a subset of three networks (a small, a medium, and a large, on the X-axis)) for increasing cache sizes (bars within each group). 
For AlexNet (the small network), Occam was able to use a single partition even with a 3 MB cache. As such, increasing the cache capacity cannot further improve performance, as  verified by our results (left-most bars). 

The partitions are completely different for different cache sizes.
In general, a larger cache results in fewer partitions as more layers fit in a partition. The new partitions can result in corner effects depending on the  spare cache capacity after Occam's partitioning. When there is abundant spare capacity, Occam   reduces its tiling overheads by fitting larger tiles. However, low spare capacity  (because of packing more layers in each partition) may result in smaller tiles and increased overhead. 
For ResNet34 (the medium network), we see  a slight drop in speedup as the cache capacity increases. Indeed, those configurations use smaller tiles and end up with slightly more tiling overheads. ResNet101 (the large network), on the other hand, improves its speedup with more cache capacity (also exhibits small corner effects). Again, this trend is closely correlated with tile size. 

\figput{csense}{}{Occam's Sensitivity to Cache Capacity}

\end{comment}

\putsubsec{fpga-results}{FPGA Results}
We use a Terasic DE2-150 FPGA development board with an Intel/Altera Cyclone IV family FPGA. 
The FPGA has approximately 150K logic elements, 820 KB of on-chip RAM  and over 300 18-bit fixed-point multipliers running at 50 MHz. The FPGA is interfaced to an external SDRAM which  offers 350 MB/s bandwidth. While the logic is adequate for our 64-lane cluster, the limited on-chip memory led us to use thinned neural networks by halving the number of channels and filters.
We use Intel's Quartus Prime (v15.0) with Qsys system builder to integrate our System Verilog implementation (not SystemC) with a soft-processor core (Nios II). Our implementation uses 99K (of the 150K available) LEs and all of the M9K RAM blocks. Like our simulation, we warm up the on-FPGA RAM with each partition's filters to capture chip residency with a single FPGA (\secref{method}). Because FPGA implementation is time-consuming, we compare only the base case and Occam. 

We implement a single cluster of 64 lanes each comprising a 
multiply-accumulate (MAC) unit (using the embedded 18-bit multipliers). Each lane holds a filter subvector (128 elements) fetched from the on-FPGA RAM for Occam (i.e., chip-resident). Each input feature map subvector is broadcast to all the lanes. Each lane computes the full input map-filter vector-vector  product to produce one output cell. 
Our implementation employs  double buffering to hide on-chip and off-chip memory latency. Our implementation uses 99K (of the 150K available) LEs and all of the M9K RAM blocks.
The  Nios II provides commands for (1) each subvector-subvector multiply and (2) fetching the filter and the input map subvectors. 
Like our software implementation, we warm up the on-FPGA RAM with each partition's filters to capture chip residency  with a single FPGA (\secref{method}). Because FPGA implementation is time-consuming, we compare only the base case and Occam. 

\tabput{thinsavings}
{\begin{tabular}{|ccccc|}
\hline
Network & AlexNet & Res34 & Res101 & Geo. Mean \\
\hline
Traffic Reduction & 7x & 31x & 43x & 21x \\
\hline
\end{tabular}
} {Traffic reduction on FPGA}

\figput{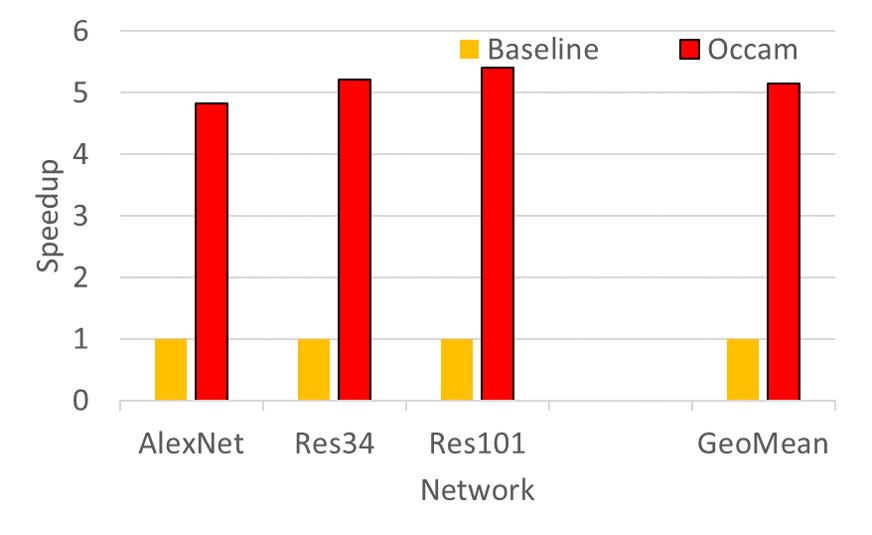}{}{Occam performance on FPGA}

\figref{fpga} shows Occam's speedups over the baseline using our FPGA implementation 

for three of our networks. 
Because the {\em baseline} FPGA implementation avoids instruction overheads and is more compute-efficient than a GPGPU, the baseline puts more pressure on memory. Consequently, 
Occam's considerable memory traffic reduction, 21x on average (\tabref{thinsavings}), results in higher speedups, 5.1x on average, with the FPGA than with a GPGPU (\figref{main}).

\putsec{related}{Related Work}

There is a plethora of work on CNNs in machine learning (ML) research~\cite{cnn-survey}. While exciting innovations in ML continue to improve CNN accuracy and efficiency, our focus is efficient execution of CNNs. \secref{intro} covers past work on CNN architectures, such as compute optimizations~\cite{snowflake, ferdman-fpga2, alexnet, quantization, moshovos-many, moshovos-cnn2, TPU}, memory optimizations~\cite{dadiannao, pudiannao, shidiannao, eie, Dally-isca17}, and 
reuse optimizations~\cite{eyeriss, fusion}. We have extensively covered Layer Fusion, the work closest to Occam.

Occam's key contribution is optimal tiling and network partitioning. Pipelining of CNN pipelines is widespread~\cite{brainwave,Kung}; our contribution is finding the pipeline partitions that optimize offchip transfers.
Tiling~\cite{tiling, tiling-lam, tiling-lam2,tiling-mckinley,tiling-book} is a well-explored area of research. 
Occam's optimal tile shape targets convolutional reuse.  Diamond Tiling~\cite{diamond-tiling} addresses tile choices in stencil computations that force pipelined start-up and induce load imbalance. Diamond-shaped tiles address this problem. Time tiling~\cite{time-tiling} enables tiling of stencil computations on periodic domains. 
While conventional tiling typically holds only the parents, Layer Fusion and Occam perform cross-layer tiling which holds some or all the ancestors. \mithuna{PolyMage~\cite{polymage} and Halide~\cite{halide} explore cross-stage tiling for image-processing pipeline stages and use heuristics to partition the pipelines. 
In contrast, Occam employs DP to find the optimal tiles.} 

We covered pipelining work related to STAP in~\secref{smbp}.

\putsec{concl}{Conclusion}
This paper targeted improving the latency of CNN inference. While 
CNNs can avoid memory latency problems via prefetching and multithreading, memory bandwidth is a problem due to the large
data. While there is enormous   data reuse,  
previous work captures only some of this reuse. {\em Full reuse} implies that the initial input image and filters are read  once from   and the final output is written once to off chip without spilling the intermediate layers' data. 
We  proposed {\em Occam} to capture full reuse via four  contributions. First, we identified the necessary  condition for full reuse. Second,  we proposed {\em dependence closure} as the sufficient condition to capture full reuse using the least on-chip memory. Third, because the on-chip memory needed for the full dependence closure is often too large,  we proposed a dynamic programming algorithm that optimally partitions a given  CNN 
to guarantee the least off-chip traffic at the partition boundaries
for a given on-chip memory capacity. 
While tiling is well-known, our contribution is determining the optimal cross-layer tiles for a given on-chip memory capacity.
Finally, because Occam's partitioning may result in unbalanced pipelines, we proposed {\em staggered asynchronous pipelining (STAP)} to improve throughput without perturbing the off-chip-transfer optimality of Occam.
Our simulations show that, on average, Occam cuts off-chip transfers by 21x  and achieves  2.06x
and 1.36x better performance, and 33\% and 24\% better energy than  the base case and Layer Fusion, respectively. Using an FPGA implementation, Occam achieves 5.1x better average performance than the base case. 
Occam's simplicity and effectiveness make it an attractive option for emerging CNN-based recognition.

\bibliographystyle{IEEEtranS}
\bibliography{ieeebst,local,cnn,gpu}

% Generated by IEEEtranS.bst, version: 1.13 (2008/09/30)
\begin{thebibliography}{10}
\providecommand{\url}[1]{#1}
\csname url@samestyle\endcsname
\providecommand{\newblock}{\relax}
\providecommand{\bibinfo}[2]{#2}
\providecommand{\BIBentrySTDinterwordspacing}{\spaceskip=0pt\relax}
\providecommand{\BIBentryALTinterwordstretchfactor}{4}
\providecommand{\BIBentryALTinterwordspacing}{\spaceskip=\fontdimen2\font plus
\BIBentryALTinterwordstretchfactor\fontdimen3\font minus
  \fontdimen4\font\relax}
\providecommand{\BIBforeignlanguage}[2]{{%
\expandafter\ifx\csname l@#1\endcsname\relax
\typeout{** WARNING: IEEEtranS.bst: No hyphenation pattern has been}%
\typeout{** loaded for the language `#1'. Using the pattern for}%
\typeout{** the default language instead.}%
\else
\language=\csname l@#1\endcsname
\fi
#2}}
\providecommand{\BIBdecl}{\relax}
\BIBdecl

\bibitem{moshovos-cnn2}
J.~Albericio \emph{et~al.}, ``Bit-pragmatic deep neural network computing,'' in
  \emph{Proceedings of the 50th Annual {IEEE/ACM} International Symposium on
  Microarchitecture, {MICRO} 2017}, pp. 382--394.

\bibitem{moshovos-many}
J.~Albericio \emph{et~al.}, ``Cnvlutin: Ineffectual-neuron-free deep neural
  network computing,'' in \emph{43rd {ACM/IEEE} Annual International Symposium
  on Computer Architecture, {ISCA} 2016,}, pp. 1--13.

\bibitem{fusion}
M.~Alwani \emph{et~al.}, ``Fused-layer cnn accelerators,'' in \emph{49th Annual
  IEEE/ACM International Symposium on Microarchitecture (MICRO)}, 2016.

\bibitem{diamond-tiling}
U.~Bondhugula \emph{et~al.}, ``Diamond tiling: Tiling techniques to maximize
  parallelism for stencil computations,'' \emph{IEEE Transactions on Parallel
  and Distributed Systems}, vol.~28, pp. 1285--1298, May 2017.

\bibitem{time-tiling}
U.~Bondhugula \emph{et~al.}, ``Tiling and optimizing time-iterated computations
  on periodic domains,'' in \emph{Proceedings of the 23rd International
  Conference on Parallel Architectures and Compilation}, ser. PACT '14.\hskip
  1em plus 0.5em minus 0.4em\relax ACM, 2014, pp. 39--50.

\bibitem{diannao}
T.~Chen \emph{et~al.}, ``Diannao: A small-footprint high-throughput accelerator
  for ubiquitous machine-learning,'' in \emph{Proceedings of the 19th
  International Conference on Architectural Support for Programming Languages
  and Operating Systems}, ser. ASPLOS '14.\hskip 1em plus 0.5em minus
  0.4em\relax New York, NY, USA: ACM, 2014, pp. 269--284.

\bibitem{eyeriss}
Y.-H. Chen \emph{et~al.}, ``14.5 eyeriss: An energy-efficient reconfigurable
  accelerator for deep convolutional neural networks,'' in \emph{2016 IEEE
  International Solid-State Circuits Conference (ISSCC)}, Jan 2016, pp.
  262--263.

\bibitem{dadiannao}
Y.~Chen \emph{et~al.}, ``Dadiannao: A machine-learning supercomputer,'' in
  \emph{Proceedings of the 47th Annual IEEE/ACM International Symposium on
  Microarchitecture}, ser. MICRO-47.\hskip 1em plus 0.5em minus 0.4em\relax
  IEEE Computer Society, 2014, pp. 609--622.

\bibitem{tiling-mckinley}
S.~Coleman and K.~S. McKinley, ``Tile size selection using cache organization
  and data layout,'' in \emph{Proceedings of the ACM SIGPLAN 1995 Conference on
  Programming Language Design and Implementation}, ser. PLDI '95.\hskip 1em
  plus 0.5em minus 0.4em\relax ACM, 1995, pp. 279--290.

\bibitem{cormen}
T.~H. Cormen \emph{et~al.}, \emph{Introduction to Algorithms, Third Edition},
  3rd~ed.\hskip 1em plus 0.5em minus 0.4em\relax The MIT Press, 2009.

\bibitem{Dally-100x}
W.~J. Dally \emph{et~al.}, ``Stream processors: Progammability and
  efficiency,'' \emph{Queue}, vol.~2, pp. 52--62, Mar. 2004.

\bibitem{shidiannao}
Z.~Du \emph{et~al.}, ``Shidiannao: Shifting vision processing closer to the
  sensor,'' in \emph{Proceedings of the 42Nd Annual International Symposium on
  Computer Architecture}, ser. ISCA '15.\hskip 1em plus 0.5em minus 0.4em\relax
  ACM, 2015, pp. 92--104.

\bibitem{brainwave}
J.~Fowers \emph{et~al.}, ``A configurable cloud-scale dnn processor for
  real-time ai,'' in \emph{Proceedings of the 45th Annual International
  Symposium on Computer Architecture}, ser. ISCA '18.\hskip 1em plus 0.5em
  minus 0.4em\relax IEEE Press, 2018, pp. 1--14.

\bibitem{snowflake}
V.~Gokhale \emph{et~al.}, ``Snowflake: An efficient hardware accelerator for
  convolutional neural networks,'' in \emph{2017 IEEE International Symposium
  on Circuits and Systems (ISCAS)}, May 2017, pp. 1--4.

\bibitem{eie}
S.~Han \emph{et~al.}, ``Eie: Efficient inference engine on compressed deep
  neural network,'' in \emph{2016 ACM/IEEE 43rd Annual International Symposium
  on Computer Architecture (ISCA)}, June 2016, pp. 243--254.

\bibitem{pipedream-arxiv}
\BIBentryALTinterwordspacing
A.~Harlap \emph{et~al.}, ``Pipedream: Fast and efficient pipeline parallel
  {DNN} training,'' \emph{CoRR}, vol. abs/1806.03377, 2018. [Online].
  Available: \url{http://arxiv.org/abs/1806.03377}
\BIBentrySTDinterwordspacing

\bibitem{pipedream-poster}
A.~Harlap \emph{et~al.}, ``Pipedream: Pipeline parallelism for {DNN}
  training,'' 2018, {SysML 2018}: Conference on Systems and Machine Learning,
  {E}xtended Abstract and Poster.

\bibitem{resnet}
K.~He \emph{et~al.}, ``Deep residual learning for image recognition,''
  \emph{CoRR}, vol. abs/1512.03385, 2015.

\bibitem{gpipe}
Y.~Huang \emph{et~al.}, ``Gpipe: Efficient training of giant neural networks
  using pipeline parallelism,'' \emph{CoRR}, vol. abs/1811.06965, 2018.

\bibitem{tiling}
F.~Irigoin and R.~Triolet, ``Supernode partitioning,'' in \emph{Proceedings of
  the 15th ACM SIGPLAN-SIGACT Symposium on Principles of Programming
  Languages}, ser. POPL '88.\hskip 1em plus 0.5em minus 0.4em\relax ACM, 1988,
  pp. 319--329.

\bibitem{layer-wise}
Z.~Jia \emph{et~al.}, ``Exploring hidden dimensions in parallelizing
  convolutional neural networks,'' \emph{CoRR}, vol. abs/1802.04924, 2018.

\bibitem{TPU}
N.~P. Jouppi \emph{et~al.}, ``In-datacenter performance analysis of a tensor
  processing unit,'' in \emph{Proceedings of the 44th Annual International
  Symposium on Computer Architecture}, ser. ISCA '17.\hskip 1em plus 0.5em
  minus 0.4em\relax ACM, 2017, pp. 1--12.

\bibitem{weird-trick}
A.~Krizhevsky, ``One weird trick for parallelizing convolutional neural
  networks,'' \emph{CoRR}, vol. abs/1404.5997, 2014.

\bibitem{alexnet}
A.~Krizhevsky \emph{et~al.}, ``Imagenet classification with deep convolutional
  neural networks,'' in \emph{Advances in Neural Information Processing Systems
  25}, F.~Pereira \emph{et~al.}, Eds.\hskip 1em plus 0.5em minus 0.4em\relax
  Curran Associates, Inc., 2012, pp. 1097--1105.

\bibitem{Kung}
H.~T. Kung \emph{et~al.}, ``Packing sparse convolutional neural networks for
  efficient systolic array implementations: Column combining under joint
  optimization,'' \emph{CoRR}, vol. abs/1811.04770, 2018.

\bibitem{tiling-lam2}
M.~D. Lam \emph{et~al.}, ``The cache performance and optimizations of blocked
  algorithms,'' in \emph{Proceedings of the Fourth International Conference on
  Architectural Support for Programming Languages and Operating Systems}, ser.
  ASPLOS IV.\hskip 1em plus 0.5em minus 0.4em\relax ACM, 1991, pp. 63--74.

\bibitem{lecun}
Y.~Lecun \emph{et~al.}, ``Gradient-based learning applied to document
  recognition,'' \emph{Proceedings of the IEEE}, vol.~86, pp. 2278--2324, Nov
  1998.

\bibitem{gpuwattch}
J.~Leng \emph{et~al.}, ``Gpuwattch: Enabling energy optimizations in gpgpus,''
  in \emph{Proceedings of the 40th Annual International Symposium on Computer
  Architecture}, ser. ISCA '13.\hskip 1em plus 0.5em minus 0.4em\relax ACM,
  2013, pp. 487--498.

\bibitem{quantization}
D.~D. Lin \emph{et~al.}, ``Fixed point quantization of deep convolutional
  networks,'' in \emph{Proceedings of the 33rd International Conference on
  International Conference on Machine Learning - Volume 48}, ser. ICML'16,
  2016, pp. 2849--2858.

\bibitem{pudiannao}
D.~Liu \emph{et~al.}, ``Pudiannao: A polyvalent machine learning accelerator,''
  in \emph{Proceedings of the Twentieth International Conference on
  Architectural Support for Programming Languages and Operating Systems}, ser.
  ASPLOS '15.\hskip 1em plus 0.5em minus 0.4em\relax ACM, 2015, pp. 369--381.

\bibitem{microbench1}
X.~Mei and X.~Chu, ``Dissecting gpu memory hierarchy through
  microbenchmarking,'' \emph{IEEE Transactions on Parallel and Distributed
  Systems}, vol.~28, pp. 72--86, Jan 2017.

\bibitem{micron}
Micron, ``Gddr5x: The next-generation graphics dram,''
  \url{https://www.micron.com/~/media/documents/products/technical-note/dram/tned02_gddr5x.pdf
  }.

\bibitem{polymage}
R.~T. Mullapudi \emph{et~al.}, ``Polymage: Automatic optimization for image
  processing pipelines,'' in \emph{Proceedings of the Twentieth International
  Conference on Architectural Support for Programming Languages and Operating
  Systems}, ser. ASPLOS '15.\hskip 1em plus 0.5em minus 0.4em\relax ACM, 2015,
  pp. 429--443.

\bibitem{Dally-isca17}
A.~Parashar \emph{et~al.}, ``Scnn: An accelerator for compressed-sparse
  convolutional neural networks,'' in \emph{Proceedings of the 44th Annual
  International Symposium on Computer Architecture}, ser. ISCA '17.\hskip 1em
  plus 0.5em minus 0.4em\relax ACM, 2017, pp. 27--40.

\bibitem{halide}
J.~Ragan-Kelley \emph{et~al.}, ``Halide: A language and compiler for optimizing
  parallelism, locality, and recomputation in image processing pipelines,'' in
  \emph{Proceedings of the 34th ACM SIGPLAN Conference on Programming Language
  Design and Implementation}, ser. PLDI '13.\hskip 1em plus 0.5em minus
  0.4em\relax ACM, 2013, pp. 519--530.

\bibitem{vggnet}
O.~Russakovsky \emph{et~al.}, ``Imagenet large scale visual recognition
  challenge,'' \emph{CoRR}, vol. abs/1409.0575, 2014.

\bibitem{ferdman-fpga2}
Y.~Shen \emph{et~al.}, ``{Escher}: A {CNN} accelerator with flexible buffering
  to minimize off-chip transfer,'' in \emph{25th IEEE International Symposium
  on Field-Programmable Custom Computing Machines ({FCCM})}, 2017.

\bibitem{hypar}
L.~Song \emph{et~al.}, ``Hypar: Towards hybrid parallelism for deep learning
  accelerator array,'' \emph{CoRR}, vol. abs/1901.02067, 2019.

\bibitem{tiling-lam}
M.~E. Wolf and M.~S. Lam, ``A data locality optimizing algorithm,'' in
  \emph{Proceedings of the ACM SIGPLAN 1991 Conference on Programming Language
  Design and Implementation}, ser. PLDI '91.\hskip 1em plus 0.5em minus
  0.4em\relax ACM, 1991, pp. 30--44.

\bibitem{tiling-book}
J.~Xue, \emph{Loop Tiling for Parallelism}.\hskip 1em plus 0.5em minus
  0.4em\relax Norwell, MA, USA: Kluwer Academic Publishers, 2000.

\bibitem{dilated}
F.~Yu and V.~Koltun, ``Multi-scale context aggregation by dilated
  convolutions,'' in \emph{International Conference on Learning Representations
  (ICLR)}, May 2016.

\bibitem{cpcie}
M.~A. Zainol and J.~L. Nunez-Yanez, ``Cpcie: A compression-enabled pcie core
  for energy and performance optimization,'' in \emph{2016 IEEE Nordic Circuits
  and Systems Conference (NORCAS)}, Nov 2016, pp. 1--6.

\bibitem{cnn-survey}
M.~D. Zeiler and R.~Fergus, \emph{Visualizing and Understanding Convolutional
  Networks}.\hskip 1em plus 0.5em minus 0.4em\relax Springer International
  Publishing, 2014, pp. 818--833.

\bibitem{cong}
C.~Zhang \emph{et~al.}, ``Optimizing fpga-based accelerator design for deep
  convolutional neural networks,'' in \emph{Proceedings of the 2015 ACM/SIGDA
  International Symposium on Field-Programmable Gate Arrays}, ser. FPGA
  '15.\hskip 1em plus 0.5em minus 0.4em\relax ACM, 2015, pp. 161--170.

\end{thebibliography}
\bstctlcite{IEEEexample:BSTcontrol}

\end{document}